\documentclass[aps,prb,twocolumn,groupedaddress]{revtex4-2}

\usepackage{longtable}
\usepackage{amstext,amsmath,amssymb,amsthm}
\usepackage{bm}
\usepackage{epsfig}
\usepackage{dcolumn}
\usepackage{graphicx, subfigure}
\usepackage{bbm}

\newcommand{\uf}{\mathbf{u}}
\newcommand{\bv}{\mathbf{b}}

\newcommand{\rnew}{\mathbf{r}}

\newcommand{\dd}{\mathrm{d}}

\begin{document}

\title{Driving forces on dislocations due to strain gradients and higher order gradients}

\author{Paulo C\'esar N. Pereira}
\email{pcnp@df.ufpe.br}
\author{S\'ergio W. S. Apolin\'ario}
\email{sergiowsa@df.ufpe.br} \affiliation{Departamento de F\'isica,
Universidade Federal de Pernambuco, 50670-901 Recife, PE, Brazil}

\begin{abstract}
Dislocations are topological defects known to be crucial in the onset of plasticity and in many properties of crystals. Classical Elasticity still fails to fully explain their dynamics under extreme conditions of high strain gradients and small scales, which can nowadays be scrutinized. In such conditions, corrections to the Volterra dislocation fields and to the Peach-Koehler force, for example, become relevant. One way to go beyond the Volterra solution is to consider other terms in the total Laurent series solution. This is the so called core field. One of its consequences is to predict a driving force on the dislocation due to background strain/stress gradients, which has also been suggested by other core energy calculations. Here we confirm its existence by presenting a direct observation of strain gradients driving edge dislocations in 2D atomistic simulations. We show that, in systems with scale invariance, the results for such core force can be used to obtain the total core energy, allowing a standard value for this energy to be compared with other classical methods of obtaining it. The force measured in our system differs from the prediction obtained by a direct core field analysis. Moreover, we found that higher order gradients of strains can also act as relevant forces.
\end{abstract}

\maketitle

\section{Introduction}
The idea of dislocation defects was first conceived mathematically \cite{volterra1907equilibre} and later applied in the context of plasticity \cite{taylor1934}, by considering the movement of defects in a periodic lattice. It soon became a vital feature of investigation in real three-dimensional (3D) crystals \cite{Frank1952,hirth1967theory,kubin2013dislocations}.
Since the bubble-raft model \cite{Bragg1947}, two-dimensional (2D) crystals have also been used as simple models to study dislocation dynamics (e.g., using colloids \cite{vanderMeer15356}, complex plasmas \cite{Nosenko2011} and vortices in superconductors \cite{Miguel2003}).

The individual dislocation movement is generally assumed to be governed by some well-known mechanisms: the Peach-Koehler (PK) driving force \cite{Peach1950} and the Peierls-Nabarro barrier \cite{Peierls1940,Nabarro1947} besides other possible motion's resistance, climb and diffusion mechanisms \cite{hirth1967theory,phillips2001crystals,bulatov2006computer,kubin2013dislocations}. These forces have been widely used to model plastic deformations in Discrete Dislocation Dynamics (DDD) simulations \cite{bulatov2006computer,kubin2013dislocations}, where the exact locations of all atoms can be ignored and one only needs to consider the dynamics of dislocation lines, in 3D, or points, in 2D. The validity of such mesoscale approach relies on the forces and mobility law that it considers.

The PK interactions between dislocations have power law behavior and the resulting dynamics has no intrinsic length scale (thus leading to a ``similitude principle" \cite{Zaiser2014}). The size effects and length scales emerging from DDD simulations \cite{El-Awady2015,Chakravarthy2011} and from rigorous theories \cite{Valdenarie2016,Groma2016} based on PK driving forces are usually associated with the obstacle and dislocation densities. They still cannot explain the full range of new plastic phenomena with technological impact observed, for instance, in micron and sub-micron scales \cite{GREER2011,kraft2010,GAO2016,Voyiadjis2017} (with a ``smaller is stronger" trend) and during shock loadings \cite{MEYERS200991,Remington2015,Zepeda-Ruiz2017,Wehrenberg2017}. Thus, several phenomenological and mechanism-based models have been developed, including corrections to the mobility law \cite{Gurrutxaga2016}, nonlocal Elasticity \cite{eringen2002nonlocal,Lazar2005} and strain gradient plasticity \cite{AIFANTIS1992,fleck1997,HUANG2004753,Fleck2015}.

The aim of our work is to broaden current knowledge about dislocation dynamics. Using atomistic simulations, we demonstrate the existence of driving forces on edge dislocations when in the presence of strain gradients. The strain deformations considered here, as in PK, are the background strain fields, i.e., the ones that were not generated by the dislocation itself but by other sources. A driving force proportional to their gradients has recently been predicted \cite{Clouet2011a,Iyer2015,Das2017} to appear as a consequence of the interaction between the core and the background strains. Nevertheless, such type of force has never been directly observed and identified. It is very smaller than PK in most situations and to identify it one needs the precise knowledge of both the strains and the resulting force on a dislocation. In the present work we design a system where these variables are completely known. Also, in systems with scale invariance, such as ours, the core energy can be obtained from the derivative with respect to the hydrostatic strain, which gives the unambiguous measurable force due to gradients of this strain.

Additional forces appear when higher order gradients in the strains are not negligible. In fact, we can make sense of the data by considering forces proportional to second or to third order derivatives. We decide to abandon the second gradient force possibility since it cannot be explained by a core energy consideration.

\section{Driving forces on a dislocation}
In this work, we consider Linear Elasticity in a purely two-dimensional crystal in the $ xy $ plane. The elastic potential energy of the displacement field $ \uf(\rnew) $ in the presence of a force density field $ \mathbf{f}(\rnew) $ is given by
\begin{equation}\label{energy}
E[\uf(\rnew)]=\int\left[\frac{1}{2}C_{ijkl}\varepsilon_{ij}(\rnew)\varepsilon_{kl}(\rnew)-f_i(\rnew)u_i(\rnew)\right]\dd^2r,
\end{equation}
where $\varepsilon_{ij}=(\partial_i u_j+\partial_j u_i)/2$ and the Einstein summation is used. Equilibrium condition $ \delta E/\delta u_i=0 $ gives $ C_{ijkl}\partial_j\partial_lu_k+f_i=0 $ and for the triangular lattice we have $ C_{ijkl}=B\delta_{ij}\delta_{kl}+\mu(\delta_{ik}\delta_{jl}+\delta_{il}\delta_{jk}-\delta_{ij}\delta_{kl}) $, where $ B $ and $ \mu $ are the bulk and shear modulus, respectively.

The Volterra solutions for an isolated dislocation at $ \rnew^d $ with Burgers vector $\mathbf{b}$ (defined as $\mathbf{b}=\oint\dd\uf$ for any counterclockwise closed curve enclosing $ \rnew^d $) assumes $ \mathbf{f}(\rnew)=0 $ and provides diverging strains at $ \rnew^d $. The usual procedure to avoid a diverging energy is to perform the integral (\ref{energy}) of Volterra strains only outside the core and then add a correction for the energy of the real crystal's dislocation core. 
Finally, considering the background strains (i.e., the total strains minus the dislocation ones) at $ \rnew^d $ and that the total energy contribution from the dislocation varies $ \delta E^{disl}=-\mathbf{f}^{disl}\cdot\delta\rnew^d $ when it moves a small distance $ \delta\rnew^d $, we obtain a configurational force given by
\begin{equation}\label{forcebasic}
\mathbf{f}^{disl}=\bm{\sigma}^{bg}\cdot\mathbf{b}\times\hat{\mathbf{z}}-\frac{\partial E^c}{\partial\varepsilon_{ij}^{bg}}\boldsymbol{\nabla}\varepsilon_{ij}^{bg}-\frac{\partial E^c}{\partial(\partial_k\varepsilon_{ij}^{bg})}\boldsymbol{\nabla}\partial_k\varepsilon_{ij}^{bg}-...,
\end{equation}
where $ \sigma_{ij}=C_{ijkl}\varepsilon_{kl} $ and $ E^c=E^c(\varepsilon_{ij}^{bg},\ \partial_k\varepsilon_{ij}^{bg},\ ...) $ is the core energy taken here as a function of the background strains and their derivatives. The first term is the well known Peach-Koehler force \cite{Peach1950,hirth1967theory,kubin2013dislocations,Lubarda2019}. Although the core properties dependence on background strains/stresses has long been studied, the second term in Eq. (\ref{forcebasic}), also termed as core-force, is a correction which has only recently been proposed \cite{Clouet2011a,Iyer2015,Das2017} but never confirmed in experiments or simulations. The other higher derivative terms in Eq. (\ref{forcebasic}) represent a direct generalization which keeps the linearity in the strains.

\begin{figure}
	\includegraphics[width=0.45\textwidth]{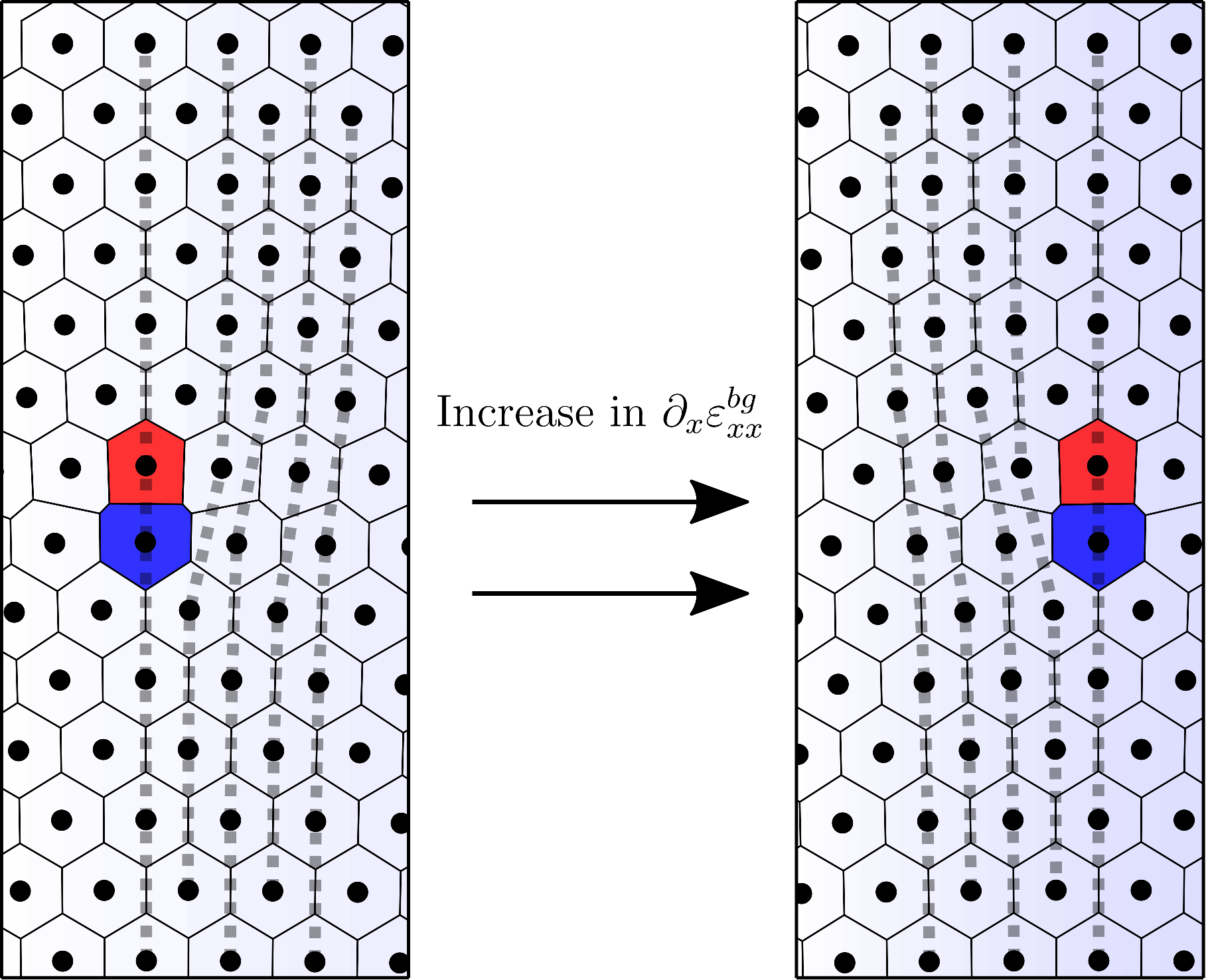}
	\caption{Sequence of configurations in a region of the simulations where we see a dislocation gliding due to the action of $ \partial_x\varepsilon_{xx}^{bg} $. In (a), the dislocation, represented by the particles with $ 5 $ (red) and $ 7 $ (blue) neighbors in the Voronoi tessellation, is in equilibrium. Here, a small PK force pointing to the left is counterbalanced by a small strain gradient force pointing to the right. The light blue shown in the background illustrates the variation in $\varepsilon_{xx}^{bg} $, increasing in the right direction. (b) As we increase the strain gradient, the dislocation moves to the right, leaving a negative resolved shear deformation behind. The vertical dotted lines are guides to the eye.}
\end{figure}

Neglecting barriers to movement, the PK force says that the dislocation can be moved if we apply a certain stress. For instance, a dislocation with $ \bv=b\hat{\mathbf{x}} $ can glide (i.e., $ \delta\rnew^d\parallel \bv $) by action of the resolved shear stress $ \sigma_{xy}^{bg}=2\mu\varepsilon_{xy}^{bg} $. In fact, the dislocation glide is itself a resolved shear deformation which can counteract the background one. In this paper, we demonstrate through Molecular Dynamics simulations that gradients and third derivatives of background strains can also glide dislocations. Fig. 1 shows how a local resolved shear deformation is induced by $ \partial_x\varepsilon_{xx}^{bg} $ through dislocation glide.

\section{Interaction of the core field with background strains}

A formal way to obtain the dislocation energy dependence on background strains is by making use of the (fictious) localized forces which can generate the core field \cite{Clouet2011a,Bacon1980}. This field is a correction to the Volterra field near the dislocation. It represents the other terms in a Laurent series solution for the elastic fields outside the core. It can be generated by considering additional point forces in the core region, i.e., $ \mathbf{f}(\rnew)=\sum_q\mathbf{F}^q\delta(\rnew-\rnew^q) $ where the positions $ \mathbf{r}^q $ are near the dislocation position $ \rnew^d $. The distances $ |\Delta \rnew^q|\equiv|\rnew^q-\rnew^d| $ are of the order of $ b $ at most. Mechanical equilibrium gives $ \sum_q \mathbf{F}_q=0 $ and, as a consequence of the reflection symmetry of the edge dislocation $ \bv=b\hat{\mathbf{x}} $, we have that for each $ q $ there exists a $ q' $ such that $ \Delta x^q=-\Delta x^{q'} $, $ F_x^q=-F_x^{q'} $, $ \Delta y^q=\Delta y^{q'} $ and $ F_y^q=F_y^{q'} $. We use these symmetries to simplify the multipole expansion of the interaction energy between the core point forces and a background deformation
\begin{eqnarray}\label{multipoleexpansion}
E^{cf-bg}=-\sum_q&& \mathbf{F}_q\cdot\uf^{bg}(\rnew_q) \nonumber\\ \approx-\big[M_1\partial_x u_x^{bg}+&&\!M_2\partial_y u_y^{bg}+M_{12}\partial_x\partial_y u_x^{bg}+M_{11}\partial_x^2 u_y^{bg}\nonumber\\
+M_{22}\partial_y^2 u_y^{bg}&&+M_{111}\partial_x^3 u_x^{bg}+M_{112}\partial_x^2\partial_y u_y^{bg}\nonumber\\
+M_{122}\partial_x\partial_y^2&& u_x^{bg}+M_{222}\partial_y^3 u_y^{bg}+\mathcal{O}(\partial^4 u^{bg})\big],
\end{eqnarray}
where the derivatives are evaluated at $ \rnew^d $, $ M_1=\sum_q F_x^q\Delta x^q $, $ M_2=\sum_q F_y^q\Delta y^q $, $ M_{12}=\sum_q F_x^q\Delta x^q\Delta y^q $, $ M_{11}=\sum_q F_y^q(\Delta x^q)^2/2 $, $ M_{22}=\sum_q F_y^q(\Delta y^q)^2/2 $, $ M_{111}=\sum_q F_x^q(\Delta x^q)^3/6 $, $ M_{112}=\sum_q F_y^q(\Delta x^q)^2\Delta y^q/2 $, $ M_{122}=\sum_q F_x^q\Delta x^q(\Delta y^q)^2/2 $ and $ M_{222}=\sum_q F_y^q(\Delta y^q)^3/6 $. Also by symmetry, a change in sign of $ b $ only changes the sign of the quadrupolar moments $ M_{12} $, $ M_{11} $ and $ M_{22} $ in Eq. (\ref{multipoleexpansion}).

We could also considerate a distribution of dislocations $\sum_q\mathbf{b}^q\delta(\rnew-\rnew^q) $ near the core with $ \sum_q \mathbf{b}_q=0 $, in addition to the point forces \cite{Clouet2011a}. Using the symmetries, this leads to the same form of Eq. (\ref{multipoleexpansion}). The resulting values of $ M_i $ are the ones that provide the appropriate core field correction for the Volterra field far from the dislocation core. For instance, the Volterra solution for the hydrostatic strain $ \boldsymbol{\nabla}\cdot\uf=\varepsilon_{ii} $ is given by
\begin{equation}\label{Volterrahs}
\varepsilon_{ii}^{V,disl}(x,y)=-\frac{(1-\beta)}{\pi}\frac{by}{x^2+y^2},
\end{equation}
where $ \beta=B/(B+\mu) $, and the core field correction is \cite{Henager2004}
\begin{equation}\label{corefieldhs}
\varepsilon_{ii}^{cf,disl}(x,y)=\frac{M^{dis}(1-\beta)}{\pi\mu}\frac{(x^2-y^2)}{(x^2+y^2)^2},
\end{equation}
where $ M^{dis}=(M_2-M_1)/2 $. Note that the Volterra and the dipolar core fields decay as $ 1/r $ and $ 1/r^2 $, respectively, while the quadrupolar and higher multipolar core fields decay more rapidly.

\begin{figure}
	\includegraphics[width=0.48\textwidth]{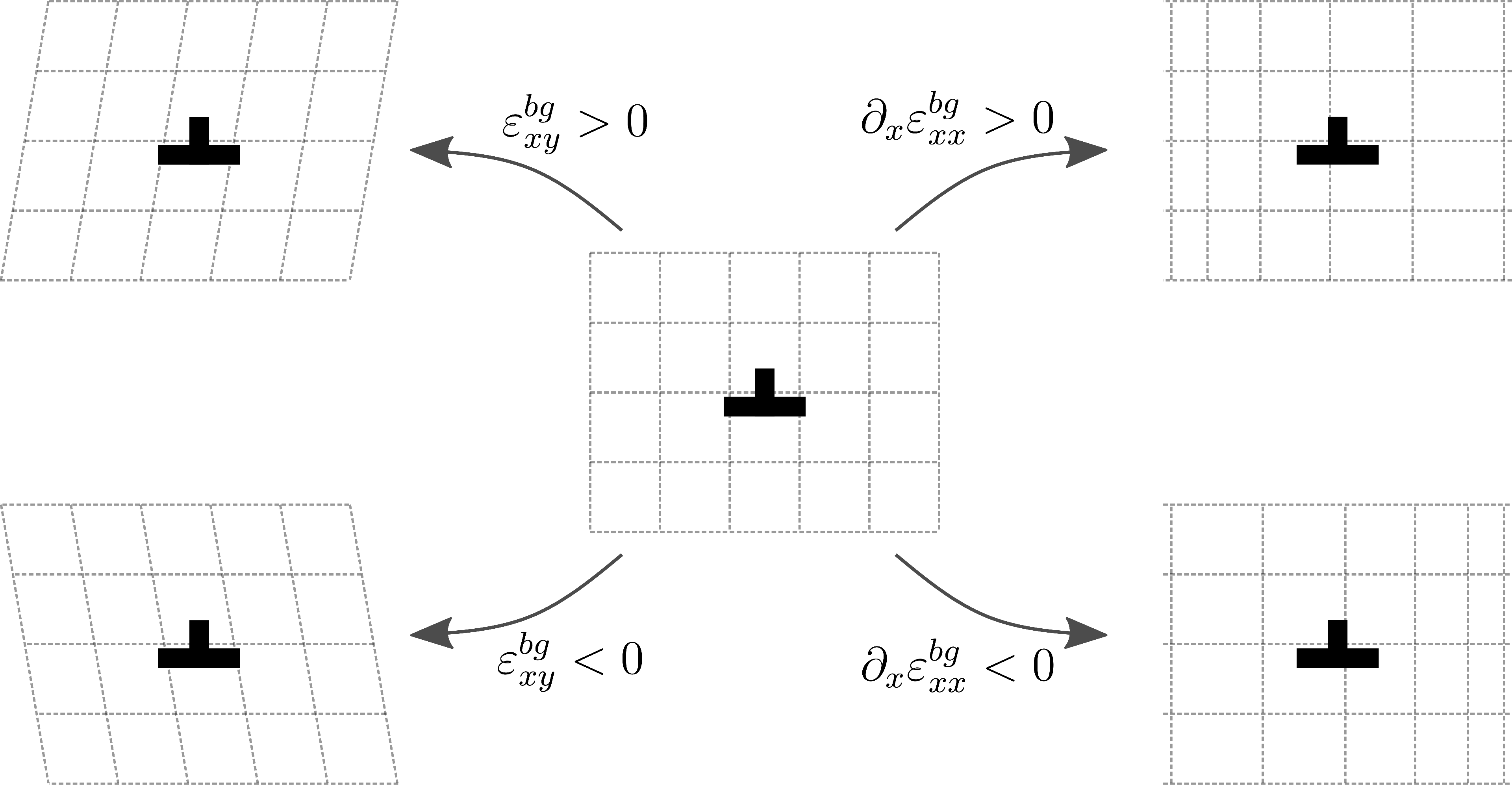}
	\caption{Illustration of the effects of $ \varepsilon_{xy}^{bg} $ and $ \partial_x\varepsilon_{xx}^{bg} $ around an edge dislocation. The reflection symmetry implies that the change in the core energy is the same for positive and negative values of these deformations. Therefore, this energy cannot depend linearly on them.}
\end{figure}

By analyzing $ \varepsilon_{ii}^{disl} $ and other dislocation deformations outside the core, we can obtain the parameters in the contribution of Eq. (\ref{multipoleexpansion}) for the interaction between the core and background strains. But the total core energy can have additional contributions for this interaction that do not relate with the core field. In fact, the Laurent series that originate this field may be divergent inside some radius, e.g., in the exact solution for the model proposed by Boleininger \emph{et al.} \cite{Boleininger2018}. Das and Gavini \cite{Das2017} used the non-singular formulation (i.e., a dislocation picture with elastic fields that do not diverge at the core) of Cai et al. \cite{Cai2006} to obtain the core-energy dependence on background strains, together with corrections obtained from the real crystal. In any case, symmetry forbids the core energy to have terms linearly proportional to $ \varepsilon_{xy}^{bg} $ and $ \partial_x\varepsilon_{xx}^{bg} $, for example, as it is show in Fig. 2. 

From now on, by ``force" we mean the configurational driving force on a dislocation, unless stated otherwise (e.g., when referring to an external force field applied on the crystal). 

Finally, using Eq. (\ref{multipoleexpansion}), the glide component of the force (\ref{forcebasic}) up to first gradient terms is given by
\begin{equation}\label{prediction}
f_x^{disl}\approx 2\mu b\varepsilon_{xy}^{bg}+M_1 \partial_x\varepsilon_{xx}^{bg}+M_2 \partial_x\varepsilon_{yy}^{bg},
\end{equation}
where we can see that the strain gradient terms do not depend on the Burgers vector. By designing conditions in which the background strains on a dislocation are known and controllable, we can not only observe if strain gradients can drive dislocation but also test if the force depends linearly with these gradients and measure its coefficients.

\section{Dislocation dipole in PBC with an external force field}

Simulations of a 2D triangular crystal with Periodic Boundary Conditions (PBC) in a rectangular box $ L_x\times L_y $ are considered. A dipole of dislocations with $ \bv=\pm a_0\hat{\mathbf{x}} $ at positions $ x=\pm d/2 $, respectively, is formed from the perfect crystal in the slip line $ y=0 $. Such dipole formation generates a discontinuity in $ u_x $ when passing from $ y<0 $ to $ y>0 $ between $ x>-d/2 $ and $ x<d/2 $, i.e., it adds $ \partial_yu_x^{(df)}=a_0\delta(y)\big[H(x+d/2)-H(x-d/2)\big] $. This results in the background homogeneous strain 
\begin{equation}\label{exybgdf}
\varepsilon_{xy}^{bg,df}=-\frac{a_0d}{2L_xL_y}
\end{equation}
due to the condition $ \int\varepsilon_{ij}\dd^2r=0 $ for the total deformation. This is a response of the system since the dislocations' movement deforms it, independently of what makes them move, as commented in the end of section II. In order to correctly identify the $x$ position of a dislocation, we only consider situations in which the two particles with number of neighbors $ \neq6 $ are aligned vertically. During the glide they move relative to each other. 

Another source of background resolved shear acting on each dislocation appears due to the interaction with the other one and its periodic images and, for each image at a distance $ \Delta\rnew $, this given by the Volterra solution
\begin{equation}\label{key}
\varepsilon_{xy}^{V,int}(\Delta x, \Delta y)=-\frac{\beta a}{2\pi}\frac{\Delta x(\Delta x^2-\Delta y^2)}{|\Delta\rnew|^4}.
\end{equation}
Therefore, the total background resolved shear contribution to the force is 
\begin{equation}\label{key}
\varepsilon_{xy}^{bg}=\varepsilon_{xy}^{bg,df}+\!\sum_{n\in\mathbb{Z},m\in\mathbb{Z}}\varepsilon_{xy}^{V,int}(d+nL_x, mL_y).
\end{equation}
We can use the fact that the sum in $ m $ is absolutely convergent and has a closed form (see Appendix A) to obtain
\begin{equation}
\varepsilon_{xy}^{bg}=-\frac{a_0d}{2L_xL_y}-\frac{\pi\beta a_0}{2L_y^2}\sum_{n\in\mathbb{Z}}(d+nL_x)\textrm{csch}^2\left[\pi\frac{(d+nL_x)}{L_y}\right],
\label{shear}
\end{equation}
where now the sum in $ n $ converges rapidly as $ |n| $ increases. Note that the dislocations' interactions do not produce net $\partial_x\varepsilon_{xx}^{bg}$ or $\partial_x\varepsilon_{yy}^{bg}$ on the slip line. 

In order to test the prediction of (\ref{prediction}), we chose to counterbalance the PK force originated from (\ref{shear}) by applying external forces which produces no $ \varepsilon_{xy}^{bg} $ but gradients in $ \varepsilon_{xx}^{bg} $ and $ \varepsilon_{yy}^{bg} $. We considered one-dimensional (1D) and radial conservative forces described by external potentials $ V_{ext}^{1D}(x)=V_{ext}^{1D}(-x) $ and $ V_{ext}^{rad}(r) $ which satisfy $ \int V_{ext}\dd^2r=0 $ on the simulation cell. For an external force field density $ \mathbf{f}(\rnew)=-\rho_0\boldsymbol{\nabla}V_{ext}(\rnew) $, where $ \rho_0=2/(\sqrt{3}a_0^2) $ is the density of the undeformed triangular crystal, the crystal's responses in the slip line $y=0$ are
\begin{equation}\label{exx1d}
\varepsilon_{xx}^{bg}(x)=\frac{\rho_0V_{ext}^{1D}(x)}{B+\mu}
\end{equation}
and $ \varepsilon_{yy}^{bg}=\varepsilon_{xy}^{bg}=0 $ in the 1D case and
\begin{equation}\label{key}
\varepsilon_{yy}^{bg}(x)=\frac{\rho_0}{(B+\mu)}\frac{1}{x^2}\int_{0}^{x}x'V_{ext}^{rad}(x')\dd x',
\end{equation}
\begin{equation}\label{key}
\varepsilon_{xx}^{bg}(x)=\frac{\rho_0V_{ext}^{rad}(x)}{B+\mu}-\varepsilon_{yy}^{bg}(x)
\end{equation}
and $ \varepsilon_{xy}^{bg}=0 $ in the radial case.

\section{Simulation details and results}

The simulations have $ N=49152 $ particles with repulsive interactions $ V_p(r)=U_0(a_0/r)^{12} $ in a cell with $ L_x=192a $ and $ L_y=256(\sqrt{3}a/2) $. Interactions like this have been used as a simple model in many situations, including dislocation studies \cite{VanSaders2018,Kapfer2015}. It is useful for our analysis due to its easiness for dislocation glide (i.e., a low Peierls-Nabarro barrier) and the bulk and shear moduli being known and with a constant ratio between them. We have $ B=21\Lambda_{12}\rho_0U_0 $ and $ \mu=(15/2)\Lambda_{12}\rho_0U_0 $, respectively, where $ \Lambda_{12}\approx6.00981 $ (see Appendix B). Power-law interactions results in the scale invariance $ E[\lambda\uf(\lambda\rnew)]=\lambda^{-12}E[\uf(\rnew)] $. The core energy, taken as an additional term in the total energy, must also satisfy this. Therefore, taking the invariance in terms of the density and this in terms of the hydrostatic strain, we have $ E^c\propto(1-\varepsilon_{ii})^6 $ and then the core energy is
\begin{equation}\label{coreenergy}
E^c_0=M^{dil}/6,
\end{equation}
where $ M^{dil}=(M_2+M_1)/2 $ and $ E^c_0 $ can depend on other background strains but its value here is for zero additional strains.

The external potentials we use are $ V_{ext}^{1D}(x)=V_0 v(x/l) $ and $ V_{ext}^{rad}(r)=V_0 v(r/l)$, where $ v(x)=x^8\big(\kappa e^{-\kappa x^{10}}-e^{-x^{10}}) $, $ \kappa=0.35 $ and we consider the cases $ l=55a_0 $, $ 57a_0 $ and $ 59a_0 $. Such external forces produce strain gradients in a region where the Peach-Koehler attraction, resulting from Eq. (\ref{shear}), is weak. Therefore, we can use small strains in the system and rely on Linear Elasticity for the theoretical analysis.

Since we do not know the dislocation mobility law completely, we consider only the situations in which the dislocations are at rest. Then, by analyzing different situations like this, we can obtain how the different contributions to the force counterbalance each other. From one equilibrium configuration to the other we use overdamped evolution for the particles in the simulations. Once in a while a small temperature is used through Brownian Dynamics \cite{Satoh} in order to avoid unstable equilibria.

\begin{figure}
	\includegraphics[width=0.47\textwidth]{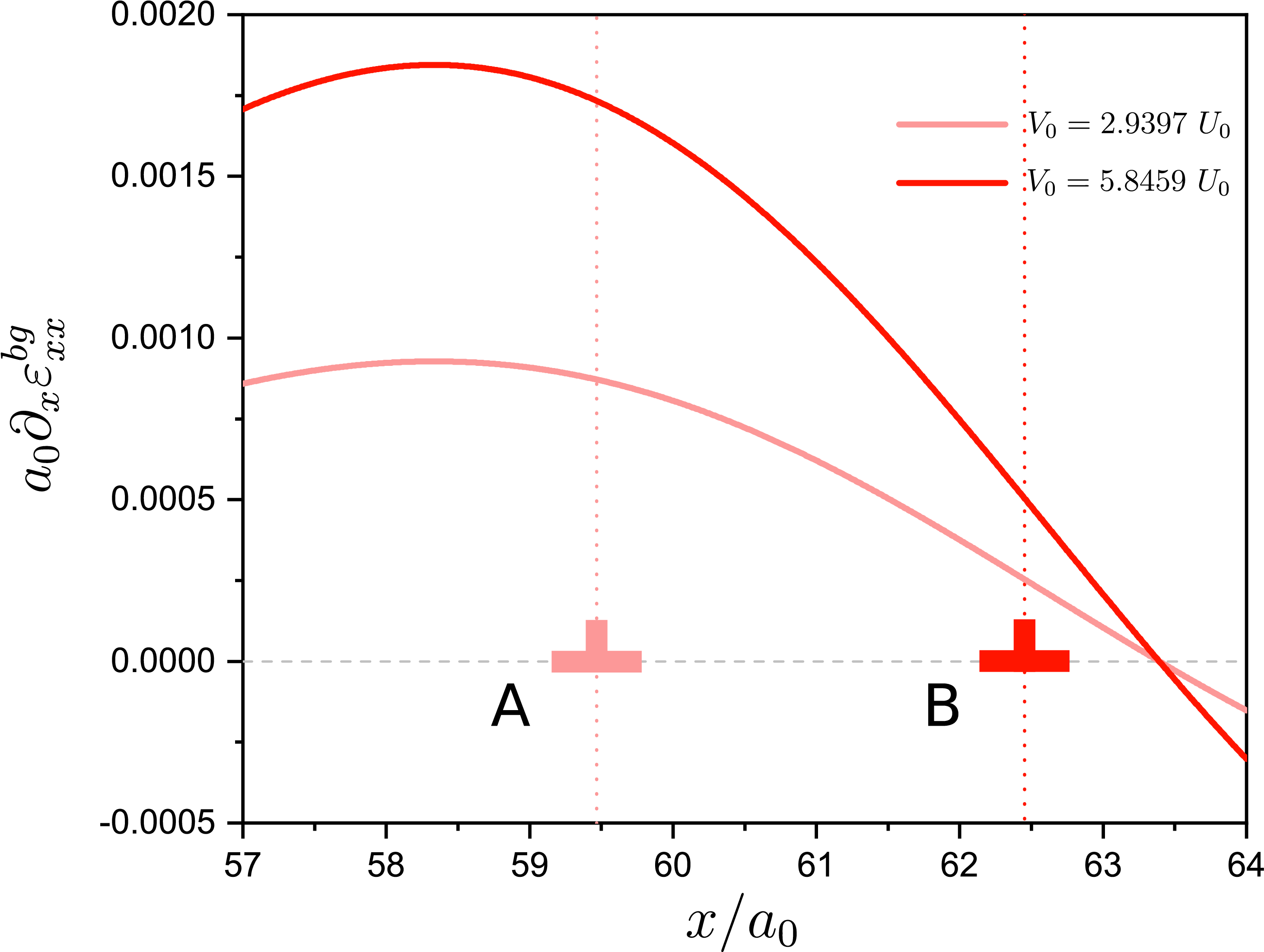}
	\caption{Plot of $ a_0\partial_x\varepsilon_{xx}^{bg}(x) $ for the system subjected to $ V_{ext}^{1D}(x)$ with $ l=55a $, $ V_0\approx2.94\varepsilon $ (light red) and $ V_0\approx5.85\varepsilon $ (dark red). For these external forces, the dislocations equilibrate, respectively, at $ x\approx\pm59.5a_0  $ and $ x\approx\pm62.5a_0 $, as shown by A and B.}
\end{figure}

The results show that strain gradients without addition of resolved shear can counteract the Peach-Koehler force and even drive the dislocation. For the system in which we apply $ V_{ext}^{1D}(x)$, Fig. 3 shows the graph of $ a_0\partial_x\varepsilon_{xx}^{bg} $ near the position at which the dislocation $ \bv=+a_0\hat{\mathbf{x}} $ is equilibrated. In this region,  $ \varepsilon_{xy}^{bg}<0 $ and, if we turn off the external force, the dislocation would move in the negative direction of $\hat{\mathbf{x}} $ and finally annihilate the other one at $ x=0 $. The figure shows that, by increasing $ V_0 $, the strain gradient increases and the dislocation moves in the right direction until it equilibrates in a new position. The same happens simultaneously for the other dislocation in the other direction.

\begin{figure}
	\includegraphics[width=0.47\textwidth]{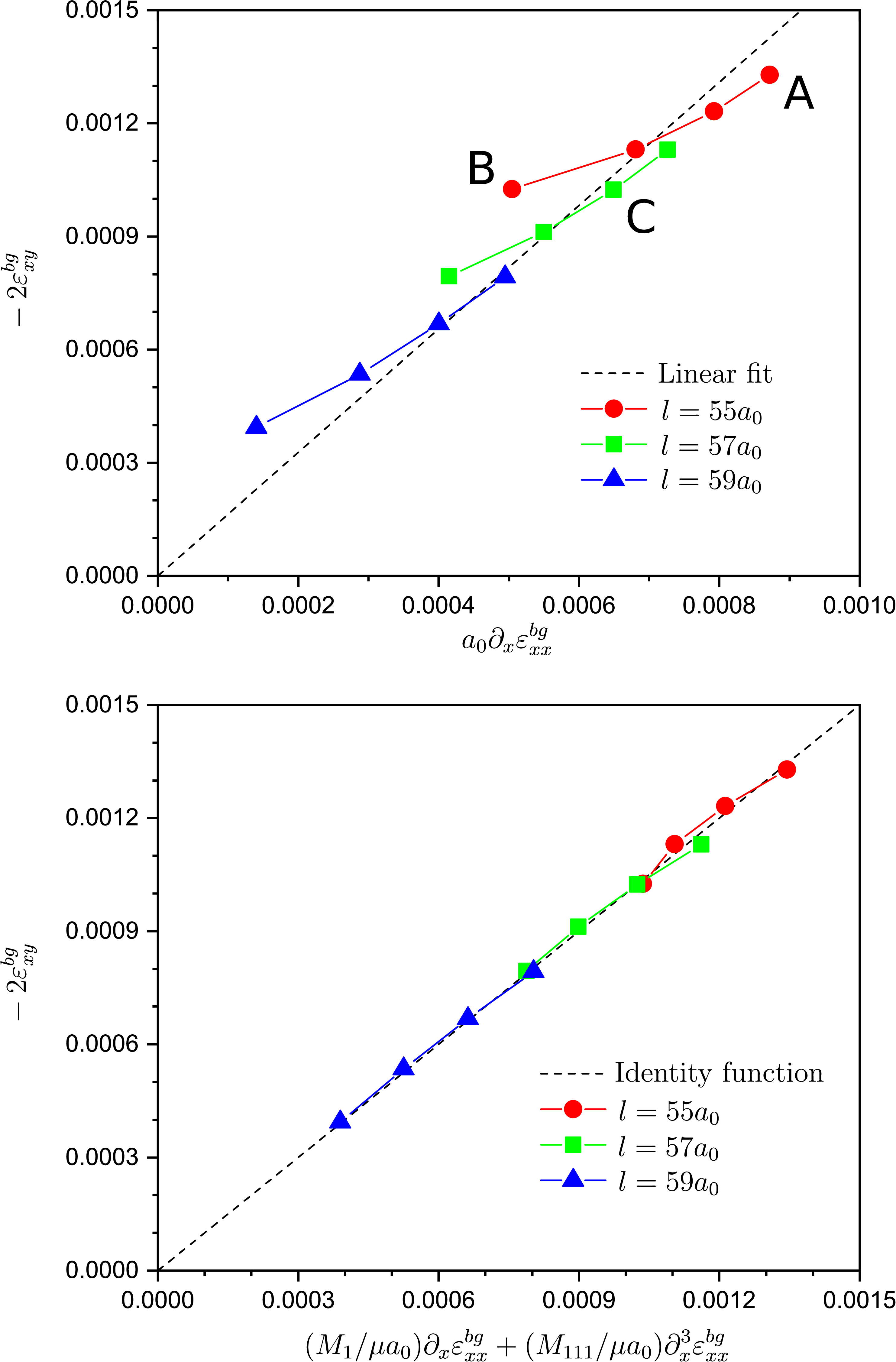}
	\caption{Relations between the background resolved shear and strain gradients on the dislocation $ \bv=+a_0\hat{\mathbf{x}} $ for equilibrium configurations in our simulations with the 1D external force field. All the deformations can be obtained from the distance between the dislocations using the equations of Sec. IV. In (a) we see that $ \partial_x\varepsilon_{xx}^{bg} $ alone cannot explain the force which is counterbalancing the PK one. By including a force due to $ \partial_x^3\varepsilon_{xx}^{bg} $, a good fit to the data is obtained, as is shown in (b).}
\end{figure}

In Fig. 4 (a) we gather the data of equilibrium distances between the dislocations for different cases of $ V_{ext}^{1D}(x)$ and use Eqs. (\ref{shear}) and (\ref{exx1d}) to plot  $ -2\varepsilon_{xy}^{bg} $ versus $ a\partial_x\varepsilon_{xx}^{bg} $, for which Eq. (\ref{prediction}) predicts a linear relation. This does not happen exactly but it shows a trend which can help us to estimate a value for $ M_1 $. A simple fit results in $M_1\approx1.64\mu a_0^2$. The cases A and B of Fig. 4 (a) are the same of Fig. 3. In the transition from A to B, the distance between the dislocations increases, $ |\varepsilon_{xy}^{bg}| $ decreases and so does $ |\partial_x\varepsilon_{xx}^{bg}| $. Furthermore, higher order derivatives also change$ |\partial_x^3\varepsilon_{xx}^{bg}| $ and they are the main difference between the situations B and C of Fig. 4 (a) besides the difference in the first derivative. By considering an additional force contribution like $ M_{11}'\partial_x^2\varepsilon_{xx}^{bg} $, we can fit well the data. But this type of force would come from a core energy which depends on $\partial_x\varepsilon_{xx}^{bg} $ and symmetry forbids this (see Fig. 2).

In order to keep a core energy description for the origin of this additional force, we consider a contribution like $ M_{111}\partial_x^3\varepsilon_{xx}^{bg} $, which comes from an octopole term in Eq. (\ref{multipoleexpansion}). This also fits well the data and we obtain $ M_1\approx2.259\mu a_0^2 $ and $ M_{111}\approx7.023\mu a_0^4 $. Fig. 4 (b) shows how the formula for the force is greatly improved by the second gradient term. Finally, we use the data for the simulations with radial forces (in which there appears $ \varepsilon_{yy}^{bg} $) and fit the equilibrium condition
\begin{eqnarray}\label{completeforce}
 f_x^{disl}&\approx&2\mu b\varepsilon_{xy}^{bg}+M_1\partial_x\varepsilon_{xx}^{bg}+M_2\partial_x\varepsilon_{yy}^{bg} \nonumber\\ &&+M_{111}\partial_x^3\varepsilon_{xx}^{bg}+M_{112}\partial_x^3\varepsilon_{yy}^{bg}=0
\end{eqnarray}
to obtain the values $ M_2\approx0.5024\mu a_0^2 $ and $ M_{112}\approx-17.57\mu a_0^4 $. The other possible background strain derivatives were not included in Eq. (\ref{completeforce}) since their values are relatively much smaller. With the values for $ M_1 $ and $ M_2 $, we can estimate the core energy given by Eq. (\ref{coreenergy}) to be $ \approx 12U_0 $.

\section{Fitting of Volterra and core field strains}

From the core field definition, we can measure $ M_1 $ and $ M_2$ by a direct analysis of the real dislocation. In order to analyze the strain fields due to the dislocation alone, we prepared the system in such a way that the dipole of dislocations has a separation $ d=L_x/2 $. A homogeneous strain which cancels the one of Eq. (\ref{exybgdf}) is then applied and another horizontal line of particles is included for the system to satisfy the PBC in the $y$ direction. We let the system equilibrate and thereafter translate it in the $x$ direction until the dislocation with $b=+a_0$ stays at $x=0$. This results in a simulation cell with $ L_x=192a $ and $ L_y=257(\sqrt{3}a/2) $ where the dislocations can equilibrate with no need of external sources of strain. Only the dislocations themselves and their images are sources of strain.

\begin{figure}
	\includegraphics[width=0.48\textwidth]{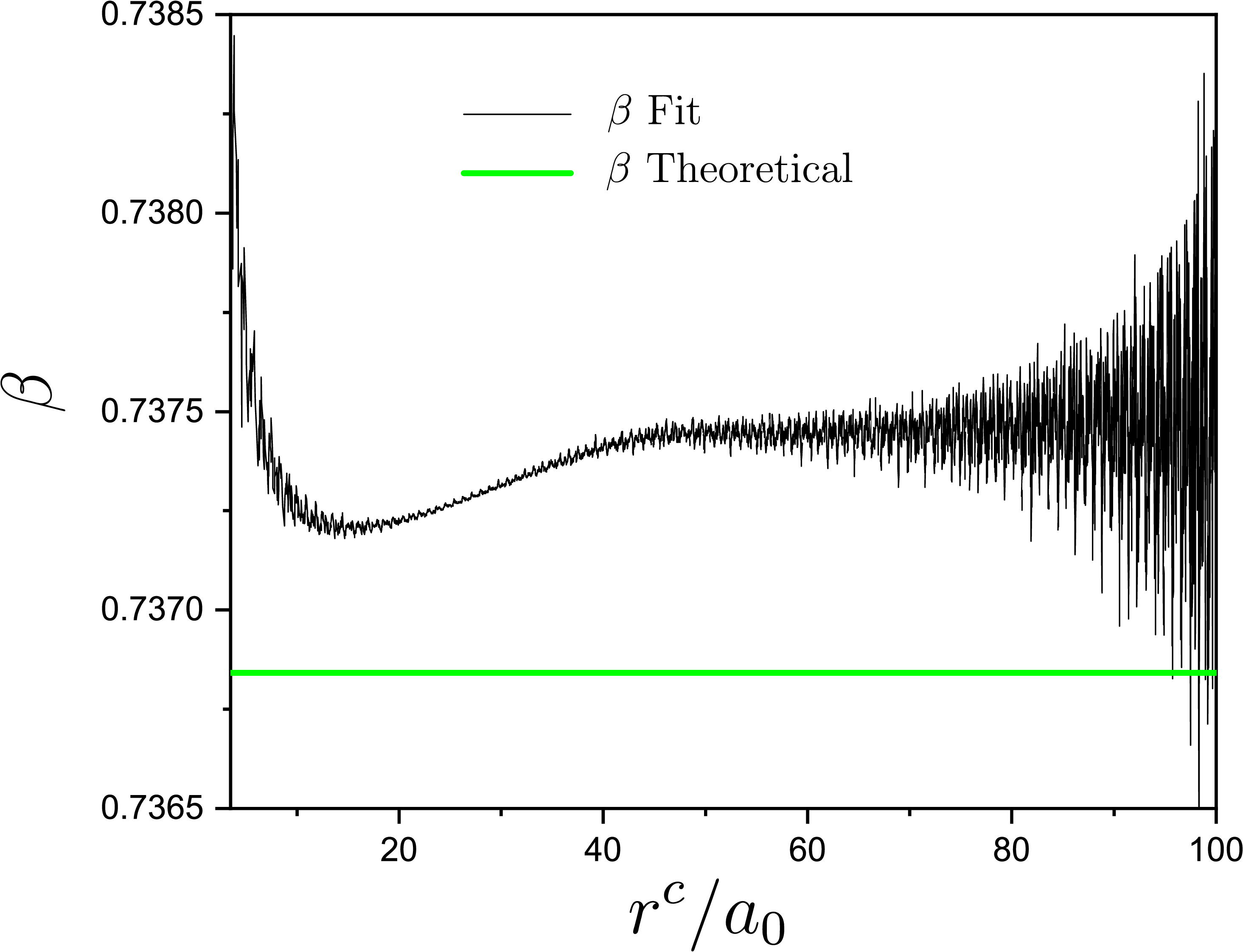}
	\caption{Values for $ \beta=B/(B+\mu) $ obtained by fitting Eq. (\ref{Volterra}) to the hydrostatic strains of the real crystal in the region outside a cutoff distance $ r^c $ from the centers of the dislocations. As  $ r^c $ increases, it fluctuates around a value near the theoretical prediction $ \beta=14/19\approx0.73684 $ (green line).}
\end{figure}

In the simulation, the strains at each particle position are estimated through the relative positions between it and its neighbors. First, we can compare the real hydrostatic strain field with the Volterra solution which, using Eq. (\ref{Volterrahs}) and a result from Appendix A, is
\begin{eqnarray}\label{Volterra}
\varepsilon_{ii}^{V}(x,y)=&&\sum_{n\in\mathbb{Z},m\in\mathbb{Z}}\big[\varepsilon_{ii}^{V,disl}(x\!+\!nL_x,\ y\!+\!mL_y)\nonumber\\
&&-\varepsilon_{ii}^{V,disl}(x\!+\!nL_x\!+\!L_x/2,\ y\!+\!mL_y)\big]\nonumber\\
=\frac{2(1\!-\!\beta)a_0}{L_x}&&\sum_{m\in\mathbb{Z}}\operatorname{Im}\!\left\lbrace\!\csc\!\left[2\pi\frac{x\!+\!i(y\!+\!mL_y)}{L_x}\right]\!\right\rbrace\!,
\end{eqnarray}
As a test, we fit Eq. (\ref{Volterra}) to the data of the actual strains in the simulation in order to obtain $ \beta $ and compare it with the theoretical value $ \beta=14/19 $. Since near the core the Volterra fields are expected to diverge from the real ones, we fitted to the data at positions farther than a cutoff distance $ r^c $ from the centers of the dislocations. Fig. 5 shows how the fitted value for $ \beta $ depends on $ r^c $. For $ r^c>d/2\equiv L_x/4 $ it fluctuates around a fixed value which is different but near the theoretical one.

\begin{figure}
	\includegraphics[width=0.47\textwidth]{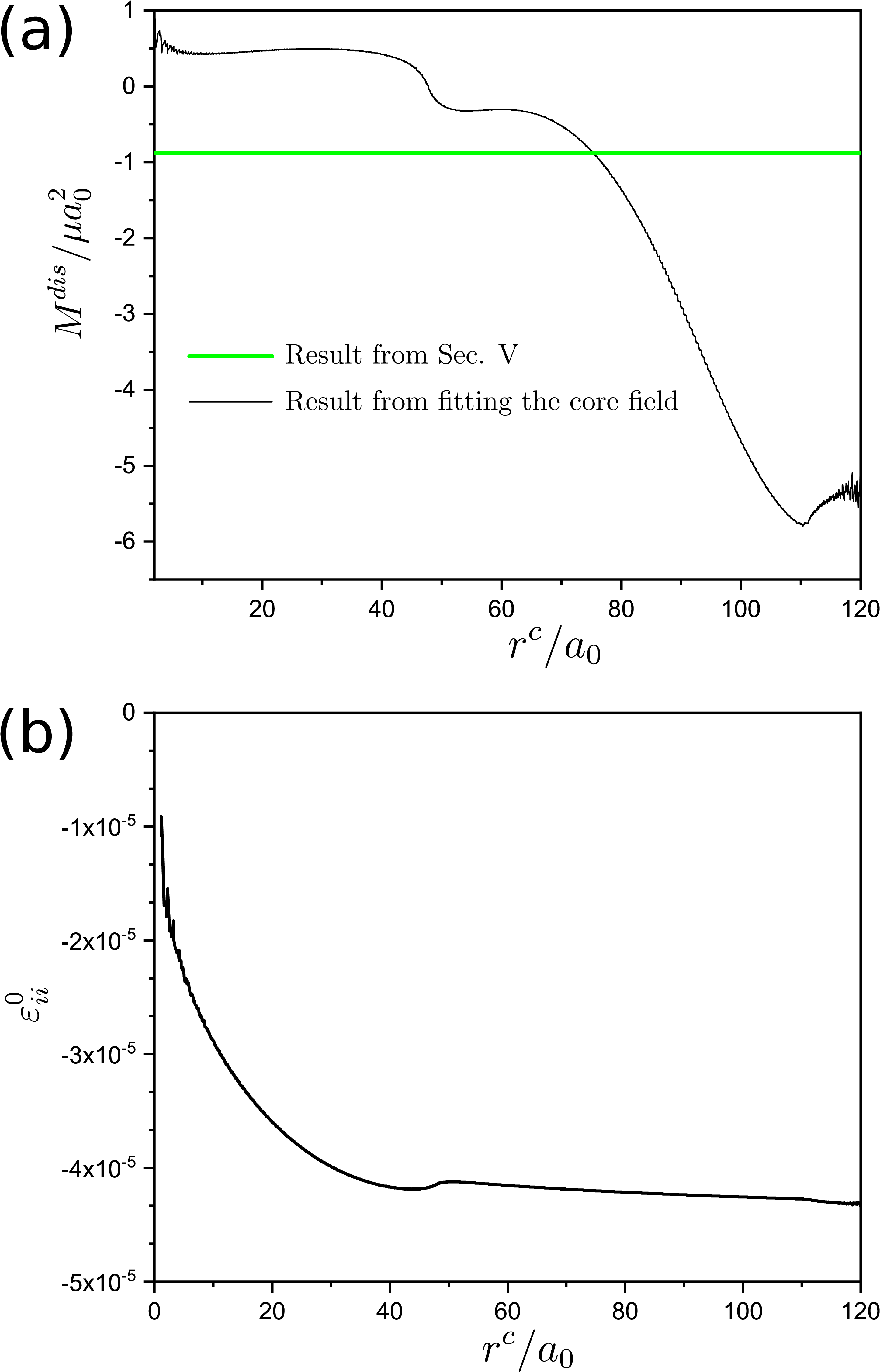}
	\caption{Values for (a) $ M^{dis} $ and (b) $ \varepsilon_{ii}^{0} $ obtained by fitting Eq. (\ref{corefield}) plus a constant $ \varepsilon_{ii}^{0} $ to the hydrostatic strains of the real crystal subtracted by Eq. (\ref{Volterra}). The fit consider points in the region outside a cutoff distance $ r^c $ from the centers of the dislocations. As  $ r^c $ increases, $ M^{dis} $ converges a value different from the one obtained in Sec. V.}
\end{figure}

Now we turn to the core field correction which, for a single dislocation, is given by Eq. (\ref{corefieldhs}). Then the core field contribution for the total hydrostatic strain is, using a result from Appendix A,
\begin{eqnarray}\label{corefield}
\varepsilon_{ii}^{cf}(x,y)=&&\sum_{n\in\mathbb{Z},m\in\mathbb{Z}}\big[\varepsilon_{ii}^{cf,disl}(x\!+\!nL_x,\ y\!+\!mL_y)\quad\nonumber\\
&&+\varepsilon_{ii}^{cf,disl}(x\!+\!nL_x\!+\!L_x/2,\ y\!+\!mL_y)\big]\nonumber\\
=\!\frac{4\pi M^{dis}(1\!-\!\beta)}{\mu L_x^2}&&\sum_{m\in\mathbb{Z}}\operatorname{Re}\!\left\lbrace\!\csc^2\!\left[2\pi\frac{x\!+\!i(y\!+\!mL_y)}{L_x}\right]\!\right\rbrace\!.
\end{eqnarray}
Moreover, the dislocations can induce a homogeneous $\varepsilon_{ii}^{0} $ in the system. Fig. 6 shows the results for $ M^{dis} $ and $\varepsilon_{ii}^{0} $ from fitting $ \varepsilon_{ii}^{cf}+\varepsilon_{ii}^{0} $ to the data of the actual strains in the simulation minus the Volterra contribution (\ref{Volterra}). As we can see, far from the core, the fit for $ M^{dis} $ converges to a value very different from the one obtained in Sec. V. From Fig. 6 (b) we infer that the dislocation core region dilates, increasing the hydrostatic strain near it (in accordance with the phenomenon of being driven to where $\varepsilon_{ii}^{bg} $ is higher) and decreasing far from the core.

\section{Conclusions}

The effects of gradients and third gradients of background strains acting as forces on edge dislocations were directly observed and quantified for the first time. This was done in a system where all the relevant strains are known and some are controllable, using 2D atomistic simulations with periodic boundary conditions.

We compare the forces with the theoretical predictions. The results are compatible with the core energy dependence on background strains as the origin for these forces. Although, the data also allows the possibility of a force proportional to second gradients, which violates a symmetry if it originates from the core behavior.

In our system, if the strain gradient force is a result solely of the core energy function, the total value for this energy can be obtained directly from the results for the force. This is due to the scale invariance in the particles interactions. The resulting core energy can be used as a standard value to be compared with other classical methods of obtaining it.

Finally, by directly analyzing the core field in the real crystal, the prediction that it gives for the difference between the coefficients of the first gradient terms in the force is very different from the one obtained by directly analyzing the force.

Strain gradient effects on the properties of the solids have gained much attention experimentally \cite{rozaliya2014strain,Tang,Goldsche2018,Wang2019} and we hope that their influence on dislocation dynamics can be seen in experiments soon.

\begin{acknowledgments}
	We thank R. M. Menezes and E. O. Lima for technical support and E. Clouet for communicating about the core field approach. We would also like to thank the Coordena\c{c}\~ao de Aperfei\c{c}oamento de Pessoal de N\'ivel Superior (CAPES) and the Conselho Nacional de Desenvolvimento Cient\'ifico e Tecnol\'ogico (CNPq) for the financial support.
\end{acknowledgments}

\appendix
\section{}

The following closed forms for infinite series are useful in our work
\begin{eqnarray}\label{key}
\sum_{n\in\mathbb{Z}} \frac{a^2-(b+n)^2}{[a^2+(b+n)^2]^2}&=&\pi^2\operatorname{Re}\big\{\!\operatorname{csch}^2[\pi (a+i b)]\big\}\\
&=&-\pi^2\operatorname{Re}\big\{\!\csc^2[\pi (b+i a)]\big\},
\end{eqnarray}
and
\begin{equation}\label{key}
\sum_{n\in\mathbb{Z}}\frac{b}{(a+n)^2+b^2}=-\pi\operatorname{Im}\big\{\!\cot[\pi (a+i b)]\big\},
\end{equation}
where $\operatorname{Re}\{z\}$ and $\operatorname{Im}\{z\}$ are the real and imaginary parts of $z$, respectively, and  $ a $ and $ b $ are real. Similar sums have been used in the context of dislocation dipoles in PBC by Zhou and colaborators \cite{Zhou2016,Zhou2017}.

\section{}
Other useful infinite sums for us are the ones which provide the bulk and shear moduli. In 2D, they are related to the pair interactions $ V_p(r) $ in the limit of zero temperature by \cite{Khrapak2018}
\begin{eqnarray}\label{key}
B=&&\lim\limits_{T \to 0}\left[2\rho_0k_BT\!-\!\frac{\pi\rho_0^2}{4}\!\int_{0}^{\infty}\!\!r^2g(r)\big[V_p'(r)\!-\!rV_p''(r)\big]\dd r\right]\nonumber\\
=&&\frac{\rho_0}{8}\sum_{\alpha}\big[-r_\alpha V_p'(r_\alpha)+r_\alpha^2V_p''(r_\alpha)\big]
\end{eqnarray}
and
\begin{eqnarray}\label{key}
\mu=&&\lim\limits_{T \to 0}\left[\rho_0k_BT\!+\!\frac{\pi\rho_0^2}{8}\!\int_{0}^{\infty}\!\!r^2g(r)\big[3V_p'(r)\!+\!rV_p''(r)\big]\dd r\right]\nonumber\\
=&&\frac{\rho_0}{8}\sum_{\alpha}\big[-r_\alpha V_p'(r_\alpha)+r_\alpha^2V_p''(r_\alpha)\big]
\end{eqnarray}
where $g(r)$ is the radial distribution function, the primes indicate derivatives and the sum is on all lattice vectors $ \rnew_\alpha $ except the origin. For the triangular lattice with power-law interactions $ V_p(r)=U_0(a_0/r)^{12} $, we have $ B=21\Lambda_{12}\rho_0U_0 $ and $ \mu=(15/2)\Lambda_{12}\rho_0U_0 $, where $ \Lambda_{12}\approx6.00981 $ is a Madelung-like lattice sum obtained by summing $ 1/(n^2+m^2+nm)^6 $ over all integer values of $n$ and $m$ except when both are zero.

\end{document}